\documentclass[12pt,a4paper,titlepage]{article}
\usepackage{fullpage}
\usepackage{amssymb,amsmath,stmaryrd}
\usepackage[mathscr]{eucal}
\usepackage{epsfig}
\usepackage{pstricks,pst-node,pst-text,pst-3d}

\newtheorem{definition}{Definition}
\newtheorem{theorem}{Theorem}
\newtheorem{proposition}{Proposition}

\newtheorem{lemma}{Lemma}

\newcommand{\axiome}[1] {{\rm{\textbf{{#1}}}}}
\newcommand{\pp}[1] { \left( {#1} \right)  }
\newcommand{\pa}[1] { \left\{ {#1} \right\}  }
\newcommand{\pr}[1] { \left[ {#1} \right]  }

\newcommand{\Sipos}{\v{S}ipo\v{s}}
\newcommand{\refe}[1] {(\ref{#1})}

\newcommand{\Q}{\mathcal{Q}}
\def \cho {\mathcal{C}}
\def \symcho {\check{\mathcal{C}}}

\newcounter{remark}

\newcounter{example}
\newenvironment{example}
{\begin{quote}\textsc{Example} \stepcounter{example} \arabic{example}:}
{\end{quote}}
\newenvironment{proof}{\medskip\noindent \bf Proof: \rm}{\hspace*{\fill}
$\blacksquare$ \newline \medskip}  

\begin{document}

\title{Bi-capacities\\ Part II: The Choquet integral\thanks{Corresponding
    author: 
    Michel GRABISCH, Universit\'e Paris I - Panth\'eon-Sorbonne.
    Tel (+33) 1-44-27-88-65, Fax (+33) 1-44-27-70-00. Email
    \texttt{Michel.Grabisch@lip6.fr}}}

\author{Michel GRABISCH\\
Universit\'e Paris I Panth\'eon-Sorbonne\\
\normalsize email \texttt{Michel.Grabisch@lip6.fr}
\and
Christophe LABREUCHE\\
Thales Research \& Technology\\
Domaine de Corbeville, 91404 Orsay Cedex, France\\
\normalsize email \texttt{Christophe.Labreuche@thalesgroup.com}}

\date{}

\maketitle

\begin{abstract}
  Bi-capacities arise as a natural generalization of capacities (or fuzzy
  measures) in a context of decision making where underlying scales are
  bipolar.  They are able to capture a wide variety of decision behaviours,
  encompassing models such as Cumulative Prospect Theory (CPT). The aim of this
  paper in two parts is to present the machinery behind bi-capacities, and thus
  remains on a rather theoretical level, although some parts are firmly rooted
  in decision theory, notably cooperative game theory. The present second part
  focuses on the definition of Choquet integral. We give several expressions of
  it, including an expression w.r.t. the M\"obius transform. This permits to
   express the Choquet integral for 2-additive bi-capacities
  w.r.t. the interaction index.

\end{abstract}

\section{Introduction}
Bi-capacities have been introduced to model in a flexible way decision
behaviours when the underlying scales are bipolar (see part I of this paper).
Specifically, let us consider a finite referential set $N=\{1,\ldots,n\}$,
which can be thought as the set of criteria (or states of nature, etc.), and
alternatives or acts, i.e., real-valued functions on $N$. Then, for $i\in N$,
$f(i)$ is the utility or score of $f$ w.r.t. to criterion $i$ (or state of
nature, etc.), which we consider to lie on a \emph{bipolar} scale, i.e., a
scale with a neutral value (usually 0) such that values above this neutral
reference point are considered as being good for the decision maker, and values
below it are considered to be bad. Assume for simplicity that the scale used
for all utilities is $[-1,1]$, with 0 as neutral value. For any disjoint
subsets $A,B$ of $N$, we denote by $v(A,B)$ the overall score or utility of the
ternary alternative $(1_A,-1_B,0_{(A\cup B)^c})$, i.e., the alternative whose
utility is equal to 1 for all criteria in $A$, to $-1$ for all criteria in $B$,
and 0 otherwise.  Assuming that $v(A,B)$ is non decreasing when $A$ increases
or $B$ decreases (in the sense of inclusion), and that
$v(\emptyset,\emptyset)=0$, $v$ viewed as a function on $\Q(N):=\{(A,B)\in
\mathcal{P}(N)\times\mathcal{P}(N) \mid A\cap B=\emptyset\}$ is called a
\emph{bi-capacity}. Its properties, as well as the M\"obius transform, the
Shapley value and the interaction have been studied at length in Part I.

\medskip

Let us consider any alternative $f\in[-1,1]^N$. The problem we address here is
to compute the overall utility of $f$ by a ``Choquet-like'' integral, which
would be defined with respect to the bicapacity $v$. Let us denote this new
Choquet integral as $\cho_v$. To achieve this properly,
we should satisfy some basic requirements. The first one is that the Choquet
integral of ternary alternatives should coincide with $v$ in the following
sense:
\[
\cho_v(1_A,-1_B,0_{(A\cup B)^c})=v(A,B)
\]
since both expressions represent the overall utility of ternary alternatives. 

The second requirement is that we should recover classical ways of computing
the overall utility of real-valued acts which are based on the (usual) Choquet
integral. These are, defining $f^+:=f\vee 0$ and $f^-:=(-f)^+$:
\begin{itemize}
\item the symmetric Choquet integral $\symcho_\nu(f):=\cho_\nu(f^+)-
  \cho_\nu(f^-)$ 
\item the asymmetric Choquet integral $\cho_\nu(f):=\cho_\nu(f^+)-
  \cho_{\overline{\nu}}(f^-) $,
where $\overline{\nu}$ indicates the conjugate capacity
\item the Cumulative Prospect Theory (CPT) model of Tversky and Kahnemann
  \cite{tvka92} $\mathrm{CPT}_{\nu_1,\nu_2}(f):=\cho_{\nu_1}(f^+) -
  \cho_{\nu_2}(f^-)$ where $\nu_1,\nu_2$ are capacities, which encompasses the
  symmetric and asymmetric Choquet integral.
\end{itemize}
Up to now, the CPT model was the most powerful model for decision making on
bipolar scales.

We will show that it is possible to propose a definition of the Choquet
integral w.r.t. bi-capacities satisfying the previous requirements. In
particular, the above mentioned models will be recovered by particular
bi-capacities, called respectively symmetric, asymmetric, bi-capacities of the
CPT type, and have already been introduced and studied in Part I of this paper.

\medskip

Section 2 introduces necessary concepts on capacities and the Choquet integral,
and on bi-capacities and bipolar capacities, the latter being a related concept
proposed by Greco \emph{et al.} \cite{grmasl02}. Section 3 gives the definition
and properties of the Choquet integral w.r.t. bi-capacities, and compares this
approach with the definition of Greco \emph{et al.} Section 4 provides the
expression of Choquet integral in terms of the M\"obius transform, finally
Section 5 focuses on 2-additive bi-capacities, and gives in this case the
expression in terms of interaction.
   
\medskip

Throughout the paper, $N:=\{1,\ldots,n\}$ denotes the finite referential set.
To avoid heavy notations, we will often omit braces and
commas to denote sets. For example, $\{i\}, \{i,j\}, \{1,2,3\}$ are denoted by 
$i, ij, 123$ respectively. The cardinality of sets will often be denoted by the
corresponding lower case, e.g., $n$ for $|N|$, $k$ for $|K|$, etc.

\section{Preliminaries}
\label{sec:prel}
We provide in this section definitions and results necessary for the sequel. For
the sake of brevity, we will refer the reader to Part I of the paper, so as to
avoid repetitions of definitions and results. However, for the convenience of the
reader we will sometimes make exceptions to this rule.

\subsection{Capacities and various transformations}
\label{sec:capa}
Let us denote by $\nu$ a capacity or a game (real-valued set function such that
$\nu(\emptyset)=0$) on $N$.
We recall from Part I that the \emph{M\"obius transform} of $\nu$ is given by
\begin{equation}
\label{eq:mob} 
m^{\nu}(A) := \sum_{B\subseteq A}(-1)^{a-b}\nu(B),
\end{equation}
with inverse relation
\begin{equation}
\nu(A) =   \sum_{B\subseteq A}m^\nu(B) \label{eq:mob3}.
\end{equation}
If $\nu$ is a capacity, it is said to be \emph{$k$-additive} if $m^{\nu}(A)=0$
for all $A\subseteq N$ such that $|A|>k$.

A convenient related transformation is the \emph{co-M\"obius transform}, defined
by
\[
\check{m}^\nu(A):=\sum_{B\supseteq N\setminus A}(-1)^{n-b}\nu(B),
\]
which is also known in Dempster-Shafer theory under the name of ``commonality
function'' \cite{sha76}, and in possibility theory \cite{dupr85b} under the
name of ``guaranted possibility measure'' (see, e.g., \cite{dupr98a}). It is
closely related with the M\"obius transform of the conjugate capacity:
\begin{equation} 
\label{eq:1.22}
\check{m}^{\bar{\nu}} (A)  =  (-1)^{|A|+1}m^\nu(A), \ \ \ \forall A\subseteq
N, A\neq\emptyset.
\end{equation}

The \emph{interaction transform} of $\nu$ is given by:
\[
I^{\nu}(A):=\sum_{B\subseteq N\setminus
A}\frac{(n-b-a)!b!}{(n-a+1)!}\Delta_A\nu(B),
\]
with $\Delta_A \nu(K) = \sum_{L\subseteq A}(-1)^{a-l}\nu(L\cup K)$, for any
$A\subseteq N$ and $K\subseteq N\setminus A$. Observe that if $\nu$ is
$k$-additive, then $I^\nu(A)=0$ for all $A\subseteq N$ such that $|A|>k$. The
following property will be used in the sequel:
\begin{equation}
\label{eq:conjint}
I^\nu(A) = (-1)^{|A|+1}I^{\overline{\nu}}(A).
\end{equation}

\subsection{Choquet integral}
\label{sec:choq}
Let $f:N\longrightarrow \mathbb{R}^+$ and $\nu$ be a capacity or a game. We
denote $f_i:=f(i)$ for simplicity. The \emph{Choquet integral} \cite{cho53} of
$f$ w.r.t. $\nu$ is defined by:
\begin{align}
\cho_\nu(f) & := \int_0^\infty \nu(\{i\mid f_i\geq\alpha\})d\alpha =
\sum_{i=1}^n 
f_{\sigma(i)}[\nu(A_{\sigma(i)}) - \nu(A_{\sigma(i+1)})] \label{eq:cho}\\
 & = \sum_{i=1}^n[f_{\sigma(i)}-f_{\sigma(i-1)}]\nu(A_{\sigma(i)}),
\end{align}
where $\sigma$ is a permutation on $N$ such that $f_{\sigma(1)}\leq\dots\leq
f_{\sigma(n)}$, $A_{\sigma(i)}:=\{\sigma(i),\dots,\sigma(n)\}$, and
$A_{\sigma(n+1)}:=\emptyset$, $f_{\sigma(0)}:=0$. A fundamental property is
that $\cho_\nu(1_A,0_{A^c})=\nu(A)$, $\forall A\subseteq N$, where
$(1_A,0_{A^c})$ is the function $f$ defined by $f_i=1$ if $i\in A$ and 0 else
(vertices of the hypercube $[0,1]^n$). It is known that the Choquet integral is
the simplest linear interpolation between vertices of the hypercube (see
\cite{gra03b}). It is also continuous w.r.t. $f_i$, for all $i\in N$. 

We consider now real valued functions $f$, and we denote by $f^+_i:=f_i\vee 0$
and $f^-_i:=(-f_i)^+$, for all $i\in N$. The \emph{asymmetric} Choquet integral
of $f$ w.r.t. $\nu$ is defined by
\begin{equation}
\cho_\nu(f):=\cho_\nu(f^+)-\cho_{\overline{\nu}}(f^-), 
\end{equation}
while the \emph{symmetric} Choquet integral (or \emph{\Sipos\ integral}) is
defined by:
\begin{equation}
\symcho_\nu(f):=\cho_\nu(f^+)-\cho_\nu(f^-) 
\end{equation}
(see, e.g., Denneberg \cite{den94}).  These two integrals are particular cases of
the Cumulative Prospect Theory model:
\begin{equation}
\label{eq:cpt}
\mathrm{CPT}_{\nu_+,\nu_-}(f):= \cho_{\nu_+}(f^+)-\cho_{\nu_-}(f^-),
\end{equation}
where $\nu_+,\nu_-$ are two capacities \cite{tvka92}. 

It is easy to get the following explicit expression of the CPT model:
\begin{multline}
\label{eq:cpt1}
\mathrm{CPT}_{\nu_+,\nu_-}(f) = \sum_{i=1}^{p-1}\big[f_{\sigma(i)} -
  f_{\sigma(i-1)}\big]\nu_-(\{\sigma(1),\ldots,\sigma(i)\}) \\
+ f_{\sigma(p)} \nu_-(\{\sigma(1),\ldots,\sigma(p)\}) + f_{\sigma(p+1)}
  \nu_+(\{\sigma(p+1),\ldots,\sigma(n)\}) \\
+ \sum_{i=p+2}^{n}\big[f_{\sigma(i)} -
  f_{\sigma(i-1)}\big]\nu_+(\{\sigma(i),\ldots,\sigma(n)\}),
\end{multline}
with $\sigma$ a permutation on $N$ such that $f_{\sigma(1)}\leq f_{\sigma(p)}
< 0 \leq f_{\sigma(p+1)}\leq f_{\sigma(n)}$. 

The expression of the CPT model with respect to the M\"obius transforms
$m^+,m^-$ and co-M\"obius transforms $\check{m}^+,\check{m}^-$ of $\nu_+,\nu_-$
is given by \cite{grlava99}:
\begin{align}
\label{eq:cptm}
\mathrm{CPT}_{\nu_+,\nu_-}(f) & =\sum_{A\subseteq N^+}m^+(A)\bigwedge_{i\in A}f_i
+ \sum_{A\subseteq N^-}m^-(A)\bigvee_{i\in A} f_i\\
\label{eq:cptco} & = \sum_{A\cap
N^+\neq\emptyset}(-1)^{|A|+1}\check{m}^+(A)\bigvee_{i\in A}f_i + \sum_{A\cap
N^-\neq\emptyset}(-1)^{|A|+1}\check{m}^-(A)\bigwedge_{i\in A}f_i
\end{align}
with $N^+:=\{i\in N\mid f_i\geq 0\}$, and $N^-:=N\setminus N^+$.

We turn now to the interaction transform. As the general expression is rather
complicated, we limit ourself to the 2-additive case. The result for the
symmetric and asymmetric integrals is the following \cite{grla00a} ($I$ denotes
the interaction transform of $\nu$).
\begin{align}
\cho_\nu(f) = & \sum_{I(ij)>0}(f_i\wedge f_j)I(ij) +
\sum_{I(ij)<0}(f_i\vee f_j)|I(ij)| + \sum_{i=1}^n
f_i(I(i)-\frac{1}{2}\sum_{j\neq i}|I(ij)|)\\
 \symcho_{\nu}(f)  =  &  
   \sum_{i,j\in N^+,I(ij)>0} (f_i \wedge f_j)I(ij)
 + \sum_{i,j\in N^-, I(ij)>0} (f_i \vee f_j) I(ij)\nonumber \\
   & + \sum_{i,j\in N^+, I(ij)<0} (f_i \vee f_j)|I(ij)|
  + \sum_{i,j\in N^-, I(ij)<0} (f_i \wedge f_j) |I(ij)| \nonumber \\
   & +  \sum_{i\in N^+} f_i \Big(\sum_{j\in N^-, \, I(ij)<0} |I(ij)|\Big)
      +  \sum_{i\in N^-} f_i \Big(\sum_{j\in N^+, \, I(ij)<0} |I(ij)|\Big)
\nonumber \\ 
   & + \sum_{i=1}^n f_i  \Big(I(i)-\frac{1}{2}\sum_{j\not= i}
|I(ij)|\Big)\label{eq:icpt} . 
\end{align}

Lastly, we give an important property of the (asymmetric) Choquet integral. Two
real-valued functions $f,g$ are \emph{comonotonic} if $f_i<f_i'$ for some
$i,i'\in N$ implies $g_i\leq g_i'$ (equivalently if $f,g$ can be made non
decreasing by the same permutation). If $f,g$ are comonotonic, then the
asymmetric Choquet integral is additive, for any game~$\nu$:
\begin{equation}
\label{eq:como}
\cho_\nu(f+g)=\cho_\nu(f)+\cho_\nu(g).
\end{equation}
For other properties of the Choquet integral, see \cite{mar00b}.

\subsection{Bi-capacities and bipolar capacities}
\label{sec:bicapa}
We denote by $\Q(N):=\{(A,B)\in \mathcal{P}(N)\times\mathcal{P}(N) \mid A\cap
B=\emptyset\}$ the set of all pairs of disjoint sets, which we endow with the
partial order defined by $(A,B)\sqsubseteq(C,D)$ iff $A\subseteq C$ and
$B\supseteq D$. Recall that a bi-capacity is an isotone mapping
$v:\mathcal{Q}(N)\longrightarrow \mathbb{R}$, with $v(\emptyset, \emptyset) =
0$. Bi-capacities of the CPT type fulfill $v(A,B) = \nu_+(A) - \nu_-(B)$, where
$\nu_+,\nu_-$ are capacities.

When $\nu_+=\nu_-$,  the bi-capacity is said to be \emph{symmetric}, and
\emph{asymmetric} when $\nu_- = \overline{\nu_+}$. 

We recall that the M\"obius transform (Part I, Section 5) of $v$ is given by
\begin{equation}
m^v(A,A') = \sum_{\substack{(B,B')\sqsubseteq(A,A')\\ B'\cap A = \emptyset}}(-1)^{|A\setminus
B|+|B'\setminus A'|}v(B,B') = \sum_{\substack{B\subseteq A\\  A'\subseteq
B'\subseteq 
A^c}}(-1)^{|A\setminus
B|+|B'\setminus A'|}v(B,B'),
\label{eq:bimob}
\end{equation}
and the inverse transform is
\begin{equation}
v(A,A') = \sum_{(B,B')\sqsubseteq(A,A')}m^v(B,B').
\label{eq:zeta}
\end{equation}
Equation (\ref{eq:zeta}) can be rewritten using bi-unanimity games $u_{(A,B)}$
defined by $u_{(A,B)}(C,D):=1$ iff $(C,D)\sqsupseteq(A,B)$, and 0 otherwise
(see (25) in Part I). This shows that, as in the classical case, the set of all
bi-unanimity games is a basis for bi-capacities:
\begin{equation}
\label{eq:una}
v(A,A') = \sum_{(B,B')\in\Q(N)}m^v(B,B')u_{(B,B')}(A,A').
\end{equation}

A bi-capacity is $k$-additive if its M\"obius transform vanishes for all
elements $(A,B)$ of $\Q(N)$ such that $|B|<n-k$.

The interaction transform of $v$ is defined by
\begin{equation}
\label{eq:biintv}
I^v_{S,T}=\sum_{K\subseteq N\setminus (S\cup T)}\frac{(n-s-t-k)!k!}{(n-s-t+1)!}
\Delta_{S,T}v(K,N\setminus(K\cup S)),
\end{equation}
for all $(S,T)\in\Q(N)$, and 
\begin{equation}
\Delta_{S,T}v(K,L) = \sum_{\substack{S'\subseteq S\\ T'\subseteq
T}}(-1)^{(s-s')+(t-t')}v(K\cup S',L\setminus T'),\quad\forall (K,L)\in
\Q(N\setminus S), L\supseteq T.
\end{equation}
The following formula will be useful.
\begin{equation}
\label{eq:biint}
I^v_{S,T} =  \sum_{(S',T')\in[(S,N\setminus(S\cup T)),(N\setminus T, \emptyset)]}\frac{1}{n-s-t-t'+1}m(S',T').
\end{equation}
In the sequel, we omit superscript $v$ for $m,I$ if no ambiguity occurs.
\medskip

On the other hand, Greco \emph{et al.} presented the notion of bipolar capacity
in \cite{grmasl02}, modelling the importance of the positive and negative parts
of the bipolar scale by pairs of numbers, and providing a different view of
bipolarity. A \emph{bipolar capacity} is a function
$\zeta:\mathcal{Q}(N)\rightarrow [0,1]\times[0,1]$ with
$\zeta(A,B)=:\pp{\zeta^+(A,B),\zeta^-(A,B)}$ such that
\begin{itemize}
\item If $A\supseteq A'$ and $B\subseteq B'$ then $\zeta^+(A,B) \geq \zeta^+(A',B')$
and $\zeta^-(A,B)\leq \zeta^-(A',B')$.
\item $\zeta^-(A,\emptyset)=0$, $\zeta^+(\emptyset,A)=0$ for any $A\subseteq N$.
\item $\zeta(N,\emptyset)=(1,0)$ and $\zeta(\emptyset,N)=(0,1)$.
\end{itemize}
In multicriteria decision making, $\zeta^+(A,B)$ can be interpreted as the
importance of coalition $A$ of criteria in the presence of $B$ for the positive
part. $\zeta^-(A,B)$ can be interpreted as the importance of coalition $B$ of
criteria in the presence of $A$ for the negative part.

This notion bears a different view of bipolarity, and can be related to the
works of Cacioppo \emph{et al.} \cite{cagabe97} on the so-called \emph{bivariate
  unipolar model}. The classical view of bipolar scales (see, e.g., Osgood
\emph{et al.} \cite{ossuta57}) amounts to consider a single axis with a central
element (the neutral value), demarcating ``good'' values from ``bad'' ones. This
is called the \emph{univariate bipolar} model. Bi-capacities are based on this
view of bipolarity, since they are valued on $[-1,1]$, which is clearly a
univariate bipolar scale with central value 0. By contrast, Cacioppo \emph{et
  al.}  consider a bipolar scale as two independent unipolar scales, one for the
positive part of the affect, and the other for the negative part of the affect
(bivariate unipolar model). The reason for such a view of bipolarity is that we
may have for an alternative both a positive and a negative feeling, and we
cannot mix them into a single resulting feeling. Clearly, bipolar capacities
rely on this view of bipolarity, since they are valued on $[0,1]^2$.

\section{The Choquet integral w.r.t. bi-capacities}

\subsection{Definition and properties}
The expression of the Choquet integral w.r.t a bi-capacity has been introduced
axiomatically in \cite{lagr02}, see also a presentation based on symmetry
considerations in \cite{grla02a}. In this section we elaborate on properties,
and relate to similar approaches. 
\begin{definition}
\label{def:bicho}
Let $v$ be a bi-capacity and $f$ be a real-valued function on $N$. The
\emph{(general) Choquet integral} of $f$ w.r.t $v$ is given by
\[
\cho_v(f) := \cho_{\nu_{N^+}}(|f|)
\] 
where $\nu_{N^+}$ is a game on $N$ defined by 
\[
\nu_{N^+}(C) := v(C\cap N^+, C\cap N^-), 
\]
and $N^+:=\{i\in N|f_i\geq 0\}$, $N^-=N\setminus N^+$. 
\end{definition}
Observe that we have $\cho_v(1_A,-1_B,0_{(A\cup B)^c}) = v(A,B)$ for any
$(A,B)\in\Q(N)$. 

Using (\ref{eq:cho}) an equivalent expression of $\cho_v(f)$ is:
\begin{equation}
\cho_v(f) =  \sum_{i=1}^n
|f_{\sigma(i)}|\Big[v(A_{\sigma(i)}\cap N^+,A_{\sigma(i)}\cap N^-)
 -v(A_{\sigma(i+1)}\cap N^+,A_{\sigma(i+1)}\cap N^-
\Big]  \label{eq:bicho}
\end{equation}
where $A_{\sigma(i)}:=\{\sigma(i),\ldots,\sigma(n)\}$, and $\sigma$ is a
permutation on $N$ so that $|f_{\sigma(1)}|\leq\cdots\leq|f_{\sigma(n)}|$.  As
it is the case for capacities, the Choquet integral w.r.t. bi-capacities is
continuous w.r.t $f_i$ for all $i\in N$. 

For the sake of clarity, let us give a numerical example. 
\begin{example}
We consider $N=\{1,2,3\}$, and the function $f$ on $N$ defined by $f(1)=-1$,
$f(2)=3$, and $f(3)=2$. Then $N^+=\{2,3\}$, $N^-=\{1\}$, so that
$\nu_{N^+}(C)=v(C\cap\{2,3\},C\cap \{1\})$. We obtain:
\begin{align*}
\cho_v(f) &  = \cho_{\nu_{N^+}}(|f|) \\
        & = |f(1)|\nu_{N^+}(N) + (|f(3)| - |f(1)|)\nu_{N^+}(\{2,3\}) + (|f(2)|
        - |f(3)|)\nu_{N^+}(\{2\})\\
        & = v(\{2,3\},\{1\}) + v(\{2,3\},\emptyset) + v(\{2\},\emptyset).
\end{align*}
\end{example}

The general Choquet integral has been axiomatized in \cite{lagr02} as follows.
In the sequel, $F_v$ denotes any functional on $\mathbb{R}^n$ defined w.r.t. a
bi-capacity $v$. We introduce the following axioms (names are those used in \cite{lagr02}).
\begin{quote}
\axiome{Monotonicity w.r.t. Bi-Capacities (MBC)}: For any bi-capacity $v$ on $\Q(N)$, $\forall
f,f'\in\mathbb{R}^n$,
\[ f_i\leq {f'}_i \:,\  \forall i\in N \ \Rightarrow F_v(f)\leq F_v(f')
\]
\end{quote}
\begin{quote}
  \axiome{Properly Weighted w.r.t. Bi-Capacities (PWBC)}: For any bi-capacity
  $v$, for any ternary vector, $F_v(1_A,-1_B,0_{(A\cup B)^c}) = v(A,B)$.
\end{quote}
\begin{quote}
  \axiome{Stable under Positive Linear transformations with positive shifts for
    Bi-Capacities and binary acts (SPLBC$^+$)}: For any bi-capacity $v$ on
  $\Q(N)$, for all $A,C\subseteq N$, $\alpha>0$, and $\beta \geq 0$,
\[  F_v\pp{(\alpha+\beta)_A,\beta_{A^c}} =
  \alpha F_v\pp{1_A,0_{A^c}}+\beta v(N,\emptyset).
\]
\end{quote}
\begin{quote}
  \axiome{Linearity w.r.t. Bi-Capacities (LBC)}: Let $A\subseteq N$, and
  real-valued functions $v,v_1,\ldots,v_p$ on $\Q(N)$ vanishing at
  $(\emptyset,\emptyset)$. If they satisfy $v(B,B') = \sum_{i=1}^p \alpha_i \:
  v_i(B,B')$ with $\alpha_1,\ldots,\alpha_p\in\mathbb{R}$, for all $(B,B')$
  such that $B\subseteq A$ and $B'\cap A=\emptyset$, then for all $f\in\Sigma_A$
\[ F_{v}(f) = \sum_{i=1}^p \alpha_i \: F_{v_i}(f),
\]
with $\Sigma_A:=\{f\in\mathbb{R}^n\mid f(i)\geq 0 \text{ iff } i\in A\}$.
\end{quote}
For $A\subseteq N$, consider the following application $\Pi_A:\mathbb{R}^n\rightarrow\mathbb{R}^n$
defined by
\[ \pp{\Pi_A(f)}_i = \left\{ \begin{array}{ll}
   f_i & \mbox{ if } i\in A \\
  -f_i & \mbox{ otherwise.} 
  \end{array} \right.
\]
By \axiome{(PWBC)}, $v(B,B')$ corresponds to the point 
$\pp{1_B,-1_{B'},0_{(B\cup B')}}$.
Define $\Pi_A\circ v(B,B')$ as the term of the bi-capacity
associated to the point 
\[ \Pi_A\pp{1_B,-1_{B'},0_{(B\cup B')^c}}
  = \pp{1_{(B\cap A)\cup (B'\setminus A)},-1_{(B\setminus A)\cup (B'\cap A)},
0_{(B\cup B')^c}}.
\]
Hence we set 
\[
\Pi_A\circ v(B,B'):= v\pp{(B\cap A)\cup (B'\setminus A), (B\setminus A)\cup
(B'\cap A)}.
\]
By symmetry arguments, it is reasonable to have $F_{\Pi_A\circ v} \pp{ \Pi_A(f)}$ being
equal to $F_v(f)$.
\begin{quote}
\axiome{Symmetry (Sym)}: For any $v:\Q(N)\rightarrow\mathbb{R}$,
we have for all $A\subseteq N$
\[ F_v(f) = F_{\Pi_A\circ v} \pp{ \Pi_A(f)} .
\]
\end{quote}
The following can be shown \cite{lagr02}.
\begin{theorem}
$\pa{F_v}_v$ satisfies \axiome{(LBC)}, \axiome{(MBC)}, \axiome{(PWBC)},
\axiome{(SPLBC$^+$)} and \axiome{(Sym)} if and only if for any bi-capacity
$v$, and for any
$N^+\subseteq N$, $f\in \Sigma_{N^+}$,
\[ F_v(f) = \cho_{\nu_{N^+}}\pp{f_{N^+}, -f_{N^-}}
\]
where $\nu_{N^+}(C):=v\pp{C\cap N^+,C\cap N^-}$, and
$\Sigma_{N^+}:=\pa{f\in\mathbb{R}^n \ , f_{N^+} \geq 0 \ , f_{N^-}< 0}$.
\label{T2}
\end{theorem}

\medskip

Let us present another way to derive the Choquet integral based on
interpolation considerations. As said in Section \ref{sec:choq}, the Choquet
integral w.r.t. capacities can be seen as the simplest linear interpolator
between vertices of the hypercube $[0,1]^n$ \cite{gra03b}, it is in fact the
Lov\'asz extension of pseudo-Boolean functions \cite{lov83}. The interpolation
is such that for any $f\in[0,1]^n$, the vertices which are used are those of
the simplex $\{x\in[0,1]^n\mid x_{\sigma(1)}\leq x_{\sigma(2)}\leq\cdots\leq
x_{\sigma(n)}\}$, where $\sigma$ is a permutation on $N$ such that
$f_{\sigma(1)}\leq f_{\sigma(2)}\leq\cdots\leq f_{\sigma(n)}$. 

\begin{figure}[htb]
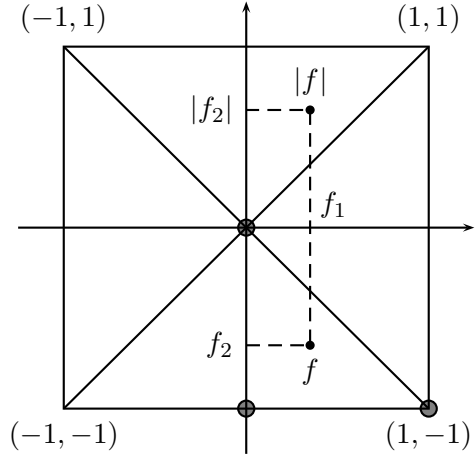

  \begin{center}
\psset{unit=1.2cm}
\pspicture(-0.5,-0.5)(4.5,4.5)
\pscircle[fillstyle=solid,fillcolor=gray](2,2){0.1}
\pscircle[fillstyle=solid,fillcolor=gray](2,0){0.1}
\pscircle[fillstyle=solid,fillcolor=gray](4,0){0.1}
\pspolygon(0,0)(4,0)(4,4)(0,4)
\psline{->}(-0.5,2)(4.5,2)
\psline{->}(2,-0.5)(2,4.5)
\psline(0,0)(4,4)
\psline(4,0)(0,4)
\psline[linestyle=dashed](2,0.7)(2.7,0.7)(2.7,3.3)(2,3.3)
\pscircle[fillstyle=solid,fillcolor=black](2.7,0.7){0.05}
\pscircle[fillstyle=solid,fillcolor=black](2.7,3.3){0.05}
\uput[90](4,4){\small $(1,1)$}
\uput[-90](0,0){\small $(-1,-1)$}
\uput[90](0,4){\small $(-1,1)$}
\uput[-90](4,0){\small $(1,-1)$}
\uput[-90](2.7,0.7){\small $f$}
\uput[90](2.7,3.3){\small $|f|$}
\uput[45](2.7,2){\small $f_1$}
\uput[180](2,0.7){\small $f_2$}
\uput[180](2,3.3){\small $|f_2|$}

\endpspicture
  \end{center}
\caption{Interpolation for the case of bi-capacities}
\label{fig:interpol3}
\end{figure}

Let us apply a similar approach for the case of bi-capacities, and call $F$ the
function we obtain by interpolation. To do this, we
examine in details the case $n=2$ (Fig. \ref{fig:interpol3}).
Let us take any point $f$ such that $f_1\geq 0$, $f_2\leq 0$ and $|f_1|\leq
|f_2|$. Then, for $|f|$ which is in the first (positive) quadrant, we already
know that the best linear interpolation is the Choquet integral. It suffices to
use the formula with the adequate vertices:
\[
F(f_1,f_2):=|f_1|F(1,-1) + (|f_2|-|f_1|)F(0,-1)
\]
This is a Choquet integral w.r.t a  game $\nu_1$ defined by:
\begin{align*}
\nu_1(\{1,2\})& = F(1,-1)\\
\nu_1(\{2\}) &= F(0,-1).
\end{align*}
Let us consider now the general case. Using $N^+=\{i\in N\mid f_i\geq 0\}$,
$N^-=N\setminus N^+$ as before, with similar considerations of symmetry, we
obtain:
\begin{multline*}
F(f) = |f_{\sigma(1)}|F(1_{N^+},-1_{N^-})\\ +  \sum_{i=2}^n(|f_{\sigma(i)}| -
|f_{\sigma(i-1)}|)F(1_{\{\sigma(i),\ldots,\sigma(n)\}\cap N^+},
-1_{\{\sigma(i),\ldots,\sigma(n)\}\cap N^-} )
\end{multline*}
where $\sigma$ is a permutation on $N$ such that
$|f_{\sigma(1)}|\leq\cdots\leq|f_{\sigma(n)}|$. This expression is the Choquet
integral of $|f|$ w.r.t. a game $\nu_{N^+}$ defined by:
\[
\nu_{N^+}(A) :=F(1_{A\cap N^+},-1_{A\cap N^-}).
\]
Recalling that $F(1_A,-1_B)=:v(A,B)$, we recover our previous Definition
\ref{def:bicho}.

\medskip

The following result shows that our definition encompasses the CPT model, and
consequently the symmetric and asymmetric Choquet integrals. 
\begin{proposition}
\label{prop:bicho}
If $v$ is of the CPT type, with $v(A,B) = \nu_+(A) - \nu_-(B)$, the Choquet
integral reduces to 
\begin{align*}
\cho_v(f) = &  \sum_{i=1}^n f^+_{\sigma(i)}\Big[\nu_+(A_{\sigma(i)}\cap N^+)-
\nu_+(A_{\sigma(i+1)}\cap N^+)] \\
        & - \sum_{i=1}^n f^-_{\sigma(i)}\Big[\nu_-(A_{\sigma(i)}\cap N^-)-
\nu_-(A_{\sigma(i+1)}\cap N^-)]\\
        = &\; \cho_{\nu_+}(f^+) - \cho_{\nu_-}(f^-).
\end{align*}
\end{proposition}
\begin{proof}
Using the definition of $v$, and splitting $N$ in $N^+,N^-$, Equation
(\ref{eq:bicho}) becomes:
\begin{multline*}
\cho_v(f) = \sum_{\sigma(i)\in N^+}f_{\sigma(i)}\Big[\nu_+(A_{\sigma(i)}\cap
N^+) - \nu_-(A_{\sigma(i)}\cap N^-)
-\nu_+(A_{\sigma(i+1)}\cap N^+) + \nu_-(A_{\sigma(i+1)}\cap N^-)\Big]\\
 - \sum_{\sigma(i)\in N^-}f^-_{\sigma(i)}\Big[\nu_+(A_{\sigma(i)}\cap
N^+) - \nu_-(A_{\sigma(i)}\cap N^-)
-\nu_+(A_{\sigma(i+1)}\cap N^+) + \nu_-(A_{\sigma(i+1)}\cap N^-)\Big].
\end{multline*}
If $\sigma(i)\in N^+$, then $A_{\sigma(i)}\cap N^-=A_{\sigma(i+1)}\cap N^-$,
and if $\sigma(i)\in N^-$, we have $A_{\sigma(i)}\cap N^+=A_{\sigma(i+1)}\cap
N^+$.  Substituting in the above expression leads to the first relation. Let us
denote by $\sigma^+,\sigma^-$ the permutations on $N$ such that $f^+,f^-$
become non decreasing. Observe that $A_{\sigma^+(i)}=A_{\sigma(i)}\cap N^+$ and
$A_{\sigma^-(i)}=A_{\sigma(i)}\cap N^-$, which proves the second relation.  
\end{proof}

Lastly, we give a property related to comonotonicity. Two real-valued functions
are said to be \emph{cosigned} if for all $i\in N$, $f_i> 0$ implies $g_i\geq
0$ and $f_i<0$ implies $g_i\leq 0$. 
\begin{proposition}
\label{prop:cocoad}
Let $f,g$ be real-valued functions on $N$. If $f,g$ are cosigned and $|f|,|g|$
are comonotonic, then for any bi-capacity $v$
\[
\cho_v(f+g) = \cho_v(f) +\cho_v(g).
\]
\end{proposition}
\begin{proof}
Let us suppose for the moment that $f_i,g_i\neq 0$, for all $i\in N$. Let us
denote $N_f:=\{i\in N\mid f_i\geq 0\}$. Due to our assumptions, we have
$N_{f+g}=N_f=N_g$, and $|f+g|=|f|+|g|$. Then, by comonotonic additivity of the
Choquet integral:
\begin{align*}
\cho_v(f+g)=\cho_{\nu_{N_{f+g}}}(|f+g|) & = \cho_{\nu_{N_{f+g}}}(|f|+|g|)\\ & =
                                \cho_{\nu_{N_f}}(|f|) + \cho_{\nu_{N_g}}(|g|)\\
                                & = \cho_v(f) + \cho_v(g).
\end{align*}
Now suppose that $f$ or $g$ are zero for some $i\in N$. Define $f',g'$ which
are identical to $f,g$, except on those $i\in N$, where we put $f'_i=\epsilon$
or $g'_i=\epsilon$. Then we are back to our previous case, where the above
result holds. Now the result holds for $f,g$ too by continuity of $\cho_v$
since we can obtain $f,g$ as limits of $f',g'$.
\end{proof}

\subsection{The Choquet integral w.r.t bipolar capacities}
Formula (\ref{eq:bicho}) is very similar to the one proposed by Greco \emph{et
  al.} \cite{grmasl02}. For $f\in\mathbb{R}^n$, let $\sigma$ be a permutation on
$N$ rearranging $|f|$ as above. Let
\[ \begin{array}{l}
   A^+_i := \pa{ \sigma(j)\: , \ j\in \pa{i,\ldots,n} \ \mbox{ such that }
   f_{\sigma(j)}\geq 0 } \\ A^-_i := \pa{ \sigma(j)\: , \ j\in \pa{i,\ldots,n} \
   \mbox{ such that } f_{\sigma(j)}\leq 0 }
  \end{array}
\]
and
\[ \begin{array}{l}
   \cho^+(f;\zeta) = \sum_{i\in N} \pp{ f_{\sigma(i)}^+ - f_{\sigma(i-1)}^+}
   \zeta^+\pp{A^+_i,A^-_i} \\ \cho^-(f;\zeta) = \sum_{i\in N} \pp{ f_{\sigma(i)}^- -
   f_{\sigma(i-1)}^-} \zeta^-\pp{A^+_i,A^-_i}
  \end{array}
\]
where $f_{\sigma(0)}:=0$, and $f^+,f^-$ defined as above. Finally the Choquet
integral w.r.t. $\zeta$ is defined by
\begin{equation}
  \cho(f;\zeta) := \cho^+(f;\zeta) - \cho^-(f;\zeta) .
\label{Ea}
\end{equation}
We illustrate the definition on the same function as in Ex. 1.
\begin{example}
Let us consider again $N=\{1,2,3\}$, and $f(1)=-1$, $f(2)=3$ and
$f(3)=2$. Applying the definition we get:
\begin{align*}
A_1^+=\{2,3\},& \quad A_1^-=\{1\}\\
A_2^+=\{2,3\},& \quad A_2^-=\emptyset\\
A_3^+=\{2\},& \quad A_3^-=\emptyset.
\end{align*}
Hence
\begin{align*}
\cho^+(f;\zeta) & = f^+(1)\zeta^+(\{2,3\},\{1\}) + (f^+(3) -
f^+(1))\zeta^+(\{2,3\},\emptyset) + (f^+(2)-f^+(3))\zeta^+(\{2\},\emptyset)\\
         & = 2\zeta^+(\{2,3\},\emptyset) +  \zeta^+(\{2\},\emptyset).\\
\cho^-(f;\zeta) & = f^-(1)\zeta^-(\{2,3\},\{1\}) + (f^-(3) -
f^-(1))\zeta^-(\{2,3\},\emptyset) + (f^-(2)-f^-(3))\zeta^-(\{2\},\emptyset)\\
        & = \zeta^-(\{2,3\},\{1\}) - \zeta^-(\{2,3\},\emptyset).
\end{align*}
Finally we obtain
\[
\cho(f;\zeta) = 2\zeta^+(\{2,3\},\emptyset) +  \zeta^+(\{2\},\emptyset) -
\zeta^-(\{2,3\},\{1\}) +\zeta^-(\{2,3\},\emptyset).
\]
\end{example}
The two definitions (Eq. (\ref{Ea}) and Def. \ref{def:bicho}) share similar
properties.  In \cite{grmasl02}, Greco \emph{et al.} mention that cosigned
comonotonic additivity (as in Prop. \ref{prop:cocoad}) holds, and even more,
that this property together with homogeneity and idempotency is sufficient to
characterize the Choquet integral w.r.t. a bipolar capacity. For further
developments along this approach of the Choquet integral, see \cite{grfi03}.

\bigskip

For $f\in\mathbb{R}^n$ for which several permutations $\sigma$ are possible, it
is easy to see that Eq. (\ref{Ea}) depends on the choice of the
permutation. This is not the case of the usual Choquet integral or the Choquet
integral w.r.t. a bi-capacity. Enforcing that the results are the same for all
permutations rearranging $|f|$ in increasing order, the following theorem shows
that we obtain some constraints on the bipolar capacity that make it reduce
to a bi-capacity.

\begin{theorem}
  The Choquet integral w.r.t. a bipolar capacity $\zeta$ is unambiguously
  defined by (\ref{Ea}) if and only if
\begin{equation}
   \forall (A,B)\in\mathcal{Q}(N) , \quad
   \zeta^+(A,B) - \zeta^-(\emptyset,B) = \zeta^+(A,\emptyset) - \zeta^-(A,B) .
\label{Eb}
\end{equation}
The bipolar capacity $\zeta$ reduces then exactly to a
bi-capacity $v$ defined by
\[ \forall (A,B)\in\mathcal{Q}(N) , \quad  v(A,B) := \zeta^+(A,B) - \zeta^-(\emptyset,B) .
\]
Moreover, the Choquet integral w.r.t. $\zeta$ is equal to $\mathcal{C}_v$.
\label{T2a}
\end{theorem}

\begin{proof}
Let us show first the necessity part.
Let $(A,B)\in\mathcal{Q}(N)$.
Consider the ternary act $f=(1_A,-1_B,0_{-A\cup B})$.
Several permutations $\sigma$ can rearrange $|f|$ in increasing order. A first one is such that
$A=\{\sigma(n-a+1)$,$\ldots$,$\sigma(n)\}$, $B=\{\sigma(n-a-b+1)$,$\ldots$,$\sigma(n-a)\}$
and $N\setminus(A\cup B)=\{\sigma(1)$,$\ldots$,$\sigma(n-a-b)\}$ where $a=|A|$ and $b=|B|$.
It is easy to see that we obtain
\[ \cho^+(f;\zeta) = \zeta^+(A,\emptyset)
\]
and
\[ \cho^-(f;\zeta) = \zeta^-(A,B)-\zeta^-(A,\emptyset) = \zeta^-(A,B)
\]
so that
\[ \cho(f;\zeta) = \zeta^+(A,\emptyset) - \zeta^-(A,B) \ .
\]
A second permutation is such that $B=\{\sigma(n-b+1)$,$\ldots$,$\sigma(n)\}$, 
$A=\{\sigma(n-a-b+1)$,$\ldots$,$\sigma(n-b)\}$
and $N\setminus(A\cup B)=\{\sigma(1)$,$\ldots$,$\sigma(n-a-b)\}$.
We get
\[ \cho^+(f;\zeta) = \zeta^+(A,B) \ \ \ , \ \ \ \cho^-(f;\zeta) = \zeta^-(\emptyset,B)
\]
and
\[ \cho(f;\zeta) = \zeta^+(A,B) - \zeta^-(\emptyset,B) \ .
\]
In order to be consistent with Eq. \refe{Ea}, the previous two expressions of
$\cho(f;\zeta)$ shall be the same. This proves Eq. \refe{Eb}.

\bigskip

Let us show now the sufficiency part.
Consider a bipolar capacity $\zeta$ satisfying \refe{Eb}.
Consider an alternative $f$.
The ambiguity of definition \refe{Ea} lies in the fact that the two terms $\cho^+(f;\zeta)$
and $\cho^-(f;\zeta)$ may depend on the choice of the permutation $\sigma$ if several permutations
fulfill $|f_{\sigma(1)}|\leq\cdots\leq|f_{\sigma(n)}|$. Henceforth, definition \refe{Ea}
is fine for sure if there is a unique permutation $\sigma$, that is to say if
all components of $f$ are different in absolute value. 

Consider $\alpha\in\pa{|f_1|,\ldots,|f_n|}$. Let $C^+=\pa{i\in N \: : \:
  f_i=\alpha}$ and $C^-=\pa{i\in N \: : \: f_i=-\alpha}$. Function
$\alpha\mapsto \cho(f;\zeta)$ with $|f_i|=\alpha$ is affine in $\alpha$ for
$\alpha$ belonging to a small interval.  Let us show that the weight of $\alpha$
in this function does not depend on the choice of the permutation $\sigma$. Let
$p-1$ be the number of components of $f$ being strictly lower than $\alpha$ in
absolute value. Let $q=p+|C^+\cup C^-|-1$. Any permutation $\sigma$ such that
$\pa{\sigma(p),\ldots,\sigma(q)}=C^+\cup C^-$ is admissible. We have four cases
\begin{itemize}
\item $\sigma(p)\in C^+$ and $\sigma(q)\in C^-$. There exist $k\in N$ and two decompositions
$C^+_1\cup \ldots \cup C^+_k$ and $C^-_1\cup \ldots \cup C^-_k$ of $C^+$ and $C^-$ respectively,
such that for any $i\in\pa{1,\ldots,k}$,
\[ C^+_i = \pa{ \sigma(p_i),\ldots,\sigma(p_i'-1)} \ \ \
   C^-_i = \pa{ \sigma(p_i'),\ldots,\sigma(p_{i+1}-1)}
\]
with $p=p_1 < p_1' < \cdots < p'_k < p_{k+1}=q+1$.
\item $\sigma(p)\in C^+$ and $\sigma(q)\in C^+$. There exist $k\in N$ and two decompositions
$C^+_1\cup \ldots \cup C^+_k$ and $C^-_1\cup \ldots \cup C^-_{k-1}$ of $C^+$ and $C^-$ respectively,
such that 
\[ \begin{array}{l}
 \forall i\in\pa{1,\ldots,k}  , \quad C^+_i = \pa{ \sigma(p_i),\ldots,\sigma(p_i'-1)} \\
 \forall i\in\pa{1,\ldots,k-1}  ,   C^-_i = \pa{ \sigma(p_i'),\ldots,\sigma(p_{i+1}-1)}
   \end{array}
\]
with $p=p_1 < p_1' < \cdots < p_k < p_k'=q+1$.
\item $\sigma(p)\in C^-$ and $\sigma(q)\in C^-$. There exist $k\in N$ and two decompositions
$C^+_1\cup \ldots \cup C^+_{k-1}$ and $C^-_1\cup \ldots \cup C^-_k$ of $C^+$ and $C^-$ respectively,
such that 
\[ \begin{array}{l}
 \forall i\in\pa{1,\ldots,k-1}  ,   C^+_i = \pa{ \sigma(p_i'),\ldots,\sigma(p_{i+1}-1)} \\
 \forall i\in\pa{1,\ldots,k}  , \quad C^-_i = \pa{ \sigma(p_i),\ldots,\sigma(p_i'-1)}
   \end{array}
\]
with $p=p_1 < p_1' < \cdots < p_k < p_k'=q+1$.
\item $\sigma(p)\in C^-$ and $\sigma(q)\in C^+$. There exist $k\in N$ and two decompositions
$C^+_1\cup \ldots \cup C^+_k$ and $C^-_1\cup \ldots \cup C^-_k$ of $C^+$ and $C^-$ respectively,
such that for any $i\in\pa{1,\ldots,k}$,
\[  C^+_i = \pa{ \sigma(p_i'),\ldots,\sigma(p_{i+1}-1)} \ \ \ 
     C^-_i = \pa{ \sigma(p_i),\ldots,\sigma(p_i'-1)}    
\]
with $p=p_1 < p_1' < \cdots < p'_k < p_{k+1}=q+1$.
\end{itemize}
These four cases are dealt with in the same way. Let us consider only the first
one. For $i\in\pa{1,\ldots,k}$, let $\mathcal{A}^+_i:=C^+_i\cup \cdots \cup
C^+_k \cup A^+_{q+1}$ and $\mathcal{A}^-_i:=C^-_i\cup \cdots \cup C^-_k \cup
A^-_{q+1}$. From the relation
\[ \cho^+(f;\zeta) = \sum_{i\in N} f_{\sigma(i)}^+ \pp{ \zeta^+\pp{A^+_i,A^-_i} 
   - \zeta^+\pp{A^+_{i+1},A^-_{i+1}}} 
\]
we get that the weight of $\alpha$ in $\cho^+(f;\zeta)$ is equal to
\[ \begin{array}{l}
  \sum_{j=1}^k \sum_{i=p_j}^{p_j'-1}
  \left[ \zeta^+\pp{\pa{\sigma(i),\ldots,\sigma(p_j'-1)}\cup\mathcal{A}^+_{j+1},\mathcal{A}^-_j} \right.\\
  \ \ \ \ \ \ \ \ \ \ \ \ \ \ \ \ \ \  \left. 
  - \zeta^+\pp{\pa{\sigma(i+1),\ldots,\sigma(p_j'-1)}\cup\mathcal{A}^+_{j+1},\mathcal{A}^-_j} \right] \\
  = \sum_{j=1}^k \pp{ \zeta^+\pp{\mathcal{A}^+_j,\mathcal{A}^-_j} 
       - \zeta^+\pp{\mathcal{A}^+_{j+1},\mathcal{A}^-_j} } \ .
  \end{array}
\]
where $\mathcal{A}^+_{k+1}:=A^+_{q+1}$.
Similarly, the weight of $\alpha$ in $\cho^-(f;\zeta)$ is equal to
\[ \sum_{j=1}^k \pp{ \zeta^-\pp{\mathcal{A}^+_{j+1},\mathcal{A}^-_j} 
       - \zeta^-\pp{\mathcal{A}^+_{j+1},\mathcal{A}^-_{j+1}} } \ ,
\]
where $\mathcal{A}^-_{k+1}:=A^-_{q+1}$.
As a consequence, the weight of $\alpha$ in $\cho(f;\zeta)$ is
\[ \begin{array}{l}
  \zeta^+\pp{\mathcal{A}^+_1,\mathcal{A}^-_1} + \zeta^-\pp{\mathcal{A}^+_{k+1},\mathcal{A}^-_{k+1}} +
\sum_{j=2}^k \pp{ \zeta^+\pp{\mathcal{A}^+_j,\mathcal{A}^-_j} + \zeta^-\pp{\mathcal{A}^+_j,\mathcal{A}^-_j}} \\
  - \sum_{j=1}^k \pp{\zeta^+\pp{\mathcal{A}^+_{j+1},\mathcal{A}^-_j} +\zeta^-\pp{\mathcal{A}^+_{j+1},\mathcal{A}^-_j} }
  \end{array}
\]
By \refe{Eb}, we have $\zeta^+(A,B)+\zeta^-(A,B)=\zeta^+(A,\emptyset)+\zeta^-(\emptyset,B)$
for any $(A,B)\in\mathcal{Q}(N)$. Hence the weight of $\alpha$ in $\cho(f;\zeta)$ is
\begin{eqnarray}
 \lefteqn{  \zeta^+\pp{A^+_p,A^-_p} + \zeta^-\pp{A^+_{q+1},A^-_{q+1}} +
\sum_{j=2}^k \pp{ \zeta^+\pp{\mathcal{A}^+_j,\emptyset} + \zeta^-\pp{\emptyset,\mathcal{A}^-_j}} } \nonumber \\
 & & - \sum_{j=1}^k \pp{\zeta^+\pp{\mathcal{A}^+_{j+1},\emptyset} +\zeta^-\pp{\emptyset,\mathcal{A}^-_j} } \nonumber \\
 & = & \pr{\zeta^+\pp{A^+_p,A^-_p}-\zeta^-\pp{\emptyset,A^-_p} } - 
       \pr{\zeta^+\pp{A^+_{q+1},\emptyset}-\zeta^-\pp{A^+_{q+1},A^-_{q+1}} } \label{Ec}
\end{eqnarray}
Since this term depends only on $p$ and $q$ but not on the decomposition of $C^+$ and $C^-$,
we conclude that the expression of $\cho(f;\zeta)$ does not depend anymore on the choice of
the permutation. Hence the {\em if} part of the proof is shown.

\bigskip

Suppose that $\zeta$ satisfies \refe{Eb}.
Let $v$ be the bi-capacity defined by 
\[ v(A,B) := \zeta^+(A,B) - \zeta^-(\emptyset,B) = \zeta^+(A,\emptyset) -
\zeta^-(A,B), \quad \forall (A,B)\in\Q(N).
\]
Since $\zeta^+(\emptyset,\emptyset)=0$ and $\zeta^-(\emptyset,\emptyset)=0$, we get
\[ \zeta^+(A,\emptyset) = v(A,\emptyset) \ \ \ , \ \ \
\zeta^-(\emptyset,B)=-v(\emptyset,B),\quad \forall A,B\subseteq N.
\]
Plugging this into previous relation, we get
\[ \zeta^+(A,B) = v(A,B) - v(\emptyset,B)  \ \ \ \mbox{ and } \ \ \ 
   \zeta^-(A,B) = v(A,\emptyset) - v(A,B) \ .
\]
This shows that $\zeta$ reduces exactly to $v$.
Finally, by \refe{Ec}, the weight of $\alpha$ in $\cho(f;\zeta)$ is
\[ v\pp{A^+_p,A^-_p} - v\pp{A^+_{q+1},A^-_{q+1}} \ .
\]
This is exactly the value of the weight of $\alpha$ in $\mathcal{C}_v(f)$.
We conclude that the Choquet integral w.r.t. $\zeta$ is equal to $\mathcal{C}_v$.
\end{proof}

The concept of bipolar
capacity reduces to bi-capacities when the Choquet integral is used. In fact, it
seems that considering the Choquet integral w.r.t. a bipolar capacity as the
pair $(\cho^+(f;\zeta),\cho^-(f;\zeta))$ would be more in the spirit of the
double unipolar model of Cacioppo \emph{et al.}

\section{Expressions of the Choquet integral with the M\"obius transform}
The following result expresses the Choquet integral in terms of the M\"obius
transform. 
\begin{proposition}
\label{prop:chomo}
For any bi-capacity $v$, any real valued function $f$ on $N$, 
\[
\cho_v(f) = \sum_{B\subseteq N} m(\emptyset,B)\Big(\bigwedge_{i\in B^c\cap N^-}
f_i\Big) + \sum_{\substack{(A,B)\in\Q(N)\\
A\neq\emptyset}}m(A,B)\Big[\Big(\bigwedge_{i\in(A\cup B)^c\cap N^-}f_i\,
+\,\bigwedge_{i\in A}f_i \Big)\vee 0\Big]
\]
with the convention $\wedge_\emptyset f_i :=0$. 
\end{proposition}
The proof is based on the following result. 
\begin{lemma}
For any real valued function $f$ on $N$, the Choquet integral of $f$ w.r.t. a
bi-unanimity game $u_{(A,B)}$ (see Section \ref{sec:bicapa} or (25) in Part I)
is given by:
\begin{equation}
\label{eq:chouna}
\cho_{u_{(A,B)}}(f)=\begin{cases}
                        0, & \text{ if } (A,B)=(\emptyset,N)\\
                        (\bigwedge_{i\in B^c}f_i)\wedge 0, & \text{ if }
                        (A,B)=(\emptyset,B)\\
                        (\bigwedge_{i\in A}f_i)\vee 0, & \text{ if } (A,B)=(A,
                        A^c), A\neq \emptyset\\ 
                        (\bigwedge_{i\in(A\cup B)^c\cap N^-}f_i+\bigwedge_{i\in
                        A}f_i)\vee 0 &, \text{ otherwise}.
                        \end{cases}
\end{equation}
\end{lemma}
\begin{proof}
Using (\ref{eq:bicho}), we get:
\[
\cho_{u_{(A,B)}}(f)=\sum_{i=1}^n
|f_{\sigma(i)}|\Big[u_{(A,B)}(A_{\sigma(i)}\cap N^+, A_{\sigma(i)}\cap N^-) -
u_{(A,B)}(A_{\sigma(i+1)}\cap N^+, A_{\sigma(i+1)}\cap N^-) \Big]
\]
with the above defined notations. For a given $i\in N$, the difference between
brackets is non zero only in the following cases: the left term is 0 and the
second one is equal to 1 (case $1_i$), or the converse (case $2_i$). Case $1_i$
happens if and only if $A_{\sigma(i+1)}\cap N^+\supset A$ and
$A_{\sigma(i+1)}\cap N^-\subseteq B$, and either $A_{\sigma(i)}\cap
N^+\not\supset A$ or $A_{\sigma(i)}\cap N^-\not\subseteq B$. Observing that the
third condition cannot occur if the first holds, this amounts to:
\[
\text{Case 1}_i \Leftrightarrow \begin{cases}
                                (\alpha_i) & \sigma(i)\in B^c\cap N^- \\
                                (\beta_i) & A_{\sigma(i)}\cap N^+\supset A \\
                                (\gamma_i) & A_{\sigma(i+1)}\cap N^-\subseteq B.
                                \end{cases}
\]
Proceeding similarly with case $2_i$, we get:
\[
\text{Case 2}_i \Leftrightarrow \begin{cases}
                                (\alpha'_i) & \sigma(i)\in A\cap N^+ \\
                                (\beta_i) & A_{\sigma(i)}\cap N^+\supset A \\
                                (\gamma_i) & A_{\sigma(i+1)}\cap N^-\subseteq B.
                                \end{cases}
\]
We remark that only the first conditions differ. 

Let us suppose that $(A,B)=(\emptyset,N)$.  Then, for any $i\in N$, neither
case $1_i$ nor case $2_i$ can happen, since conditions $(\alpha_i)$,
$(\alpha'_i)$ cannot hold. Hence $\cho_{u_{(\emptyset,N)}}(f)=0$. This proves
the first line in (\ref{eq:chouna}).

Let us suppose $(A,B)=(\emptyset,B)$, $B\neq N$. Then for any $i\in N$, case
$2_i$ can never occur since condition $(\alpha'_i)$ is not fulfilled, and
condition $(\beta_i)$ is always fulfilled. If $B^c\cap N^-\neq\emptyset$, there
is at least one $i$ such that both $(\alpha_i)$ and $(\gamma_i)$ are fulfilled,
namely $\sigma(i)\in B^c\cap N^-$ such that $|f_{\sigma(i)}|$ is maximum (this
forces $\sigma(i+1)$ to be either in $N^+$ or in $B$). Hence we get
$\cho_{u_{(\emptyset,B)}}(f) = \wedge_{i\in B^c}f_i$ under $B^c\cap
N^-\neq\emptyset$, and 0 otherwise. This proves the second line of
(\ref{eq:chouna}).

Let us suppose $(A,B)=(A, A^c)$. If condition $\beta_i$ holds for some $i$,
then we have $A\subseteq N^+$. For any $i\in N$, case $1_i$ cannot occur since
$\sigma(i)\in A\cap N^-$ by condition $\alpha_i$, so that by condition
$\beta_i$, $\sigma(i)\in N^+$, which contradicts condition $\alpha_i$. Any
$\sigma(i)\in A$ fulfills conditions $\alpha'_i$ and $\gamma_i$, while
fulfilling also condition $\beta_i$ imposes to choose $\sigma(i)\in A$ such
that $f_{\sigma(i)}$ is minimum. Hence, under these conditions we get
$\cho_{u_{(A, A^c)}}(f) = \wedge_{i\in A}f_i$. This proves the third line of
(\ref{eq:chouna}).

Let us consider the general case, with $A\neq\emptyset, B^c$. Let us suppose
that case $1_i$ occurs for some $i\in N$. Then necessarily, $A\subseteq N^+$ by
condition $\beta_i$, $\sigma(i)\in(A\cup B)^c\cap N^-\neq\emptyset$ (conditions
$\alpha_i$, $\beta_i$), and moreover $\sigma(i)$ is such that $|f_{\sigma(i)}|$
is maximum on $(A\cup B)^c\cap N^-$ and
\begin{equation}
\label{eq:cond}
|f_{\sigma(i)}|<\wedge_{j\in A}f_{\sigma(j)}
\end{equation}
(conditions $\beta_i$, $\gamma_i$). Under the assumption that case $1_i$ holds,
let us show that case $2_j$ must occur for some $j$. Since $\emptyset\neq
A\subseteq N^+$, condition $\alpha'_j$ holds for some $j$. For any such $j$,
$j>i$ by (\ref{eq:cond}). Now, condition $\beta_j$ imposes to choose $j$ such
that $f_{\sigma(j)}$ is minimum in $A\cap N^+$. $j$ being such defined, let us
show that $\gamma_j$ holds. Suppose it is not the case. This means that there
exists $j'$ such that $\sigma(j')\in A_{\sigma(j+1)}\cap N^-$ and
$\sigma(j')\not\in B$. Since $A\subseteq N^+$, this means that $\sigma(j')\in
(A\cup B)^c\cap N^-$. However, $i<j<j'$, so that
$|f_{\sigma(j')}|>|f_{\sigma(i)}|$, which contradicts the definition of
$\sigma(i)$. The reverse result can be shown as well, so that either both cases
hold or both fail to hold. If both hold, $\cho_{u_{(A,B)}}(f)=\wedge_{i\in
(A\cup B)^c\cap N^-}f_i + \wedge_{i\in A} f_i$, which is $\geq 0$ by
(\ref{eq:cond}). This proves the last line of (\ref{eq:chouna}).
\end{proof}

Using the above lemma, the proof of Prop. \ref{prop:chomo} is immediate
considering Eq. (\ref{eq:una}) and convention $\wedge_\emptyset=0$. 

We can recover the result for the CPT model (\ref{eq:cptm}) and
(\ref{eq:cptco}), using Prop. 1 of Part I, and (\ref{eq:1.22}). Recalling
the convention $\wedge_\emptyset f_i :=0$, we get:
\begin{align*}
\cho_v(f) & =  \sum_{B\subsetneq N}m^{\overline{\nu_-}}(B^c)\bigwedge_{B^c\cap
  N^-} f_i + \sum_{A\neq \emptyset}m^{\nu_+}(A)\Big(\bigwedge_{i\in A}f_i\vee
  0\Big)\\
 & = \sum_{B\cap N^-\neq\emptyset}(-1)^{b+1}\check{m}^{\nu_-}(B)\bigwedge_{i\in B} f_i +
  \sum_{A\subseteq N^+}m^{\nu_+}(A)\bigwedge_{i\in A}f_i
\end{align*}
after some simplifications.

We illustrate the expression w.r.t. the M\"obius transform taking again the
same example. 
\begin{example}
We consider as before $N=\{1,2,3\}$ and $f(1)=-1$, $f(2)=3$ and $f(3)=2$. Since
$N^-=\{1\}$, among the 8 terms in $m(\emptyset,B)$, only those such that
$1\not\in B$ are non zero. Among the 19 terms in $m(A,B)$, $A\neq\emptyset$,
those such that $1\in A$ are zero. Hence it remains
\begin{align*}
\cho_v(f) & = [m(\emptyset,\emptyset) + m(\emptyset,2) + m(\emptyset,3) +
m(\emptyset,\{2,3\})]f(1)\\
         &+ [m(2,\emptyset)+m(2,3)][(f(1)+f(2))\vee 0] +
[m(3,\emptyset)+m(3,2)][(f(1)+f(3))\vee 0] \\
         & +m(\{2,3\},\emptyset)[(f(1)+(f(2)\wedge f(3)))\vee 0]  \\
        & +[m(2,1)+m(2,\{1,3\})]f(2) + +[m(3,1)+m(3,\{1,2\})]f(3) +
m(\{2,3\},1)[f(2)\wedge f(3)] \\ 
        & = -[m(\emptyset,\emptyset) + m(\emptyset,2) + m(\emptyset,3) +
m(\emptyset,\{2,3\})] \\
       & + 2[m(2,\emptyset)+m(2,3)] + m(3,\emptyset)+m(3,2) +
m(\{2,3\},\emptyset) + 3[m(2,1)+m(2,\{1,3\})]\\
      &+ 2 [m(3,1)+m(3,\{1,2\}) + m(\{2,3\},1)]. 
\end{align*}
\end{example}

\section{The case of 2-additive bi-capacities}
\label{sec:2add}
Let us consider a 2-additive bi-capacity $v$. The M\"obius transform is non zero
for any $(A,B)$ such that $|B|\geq n-2$, i.e., for elements $(\emptyset,i^c)$,
$(\emptyset,\{i,j\}^c)$, $(i, i^c)$, and $(\{i,j\},\{i,j\}^c)$ (denoted
$(ij,(ij)^c)$ for short), for any $i,j\in N$.

We express it w.r.t. the interaction $I$. For convenience, we use the notation
$I_{S,T}$. 
\begin{proposition}
\label{prop:2mI}
For any 2-additive bi-capacity, we have:
\begin{align}
m(ij,(ij)^c) &= I_{ij,\emptyset}\\
m(\emptyset,(ij)^c) &= I_{\emptyset,ij}\\
m(i,(ij)^c) &= I_{i,j}\\
m(i,i^c) &= I_{i,\emptyset} - \frac{1}{2}\sum_{j\neq i}[I_{i,j} + I_{ij,\emptyset}]\\
m(\emptyset,i^c) &= I_{\emptyset,i} -\frac{1}{2}\sum_{j\neq
  i}[I_{j,i}+I_{\emptyset, ij}]  
\end{align}
\end{proposition}
\begin{proof}
The proof comes readily from Prop. 6 in Part I for the three first ones,
and Eq.~(\ref{eq:biint}) for the last two. We just detail the last two. We
have:
\[
I_{i,\emptyset} = m(i,i^c) + \sum_{j\neq i}\frac{1}{2}[m(i,(ij)^c) + m(ij,(ij)^c)]. 
\]  
Replacing $m(i,(ij)^c)$ and $m(ij,(ij)^c)$ by their expression in terms of $I$
gives the result. Similarly, the last equation comes from
\[
I_{\emptyset,i} = m(\emptyset,i^c) + \sum_{j\neq i}\frac{1}{2}[m(j,(ij)^c) + m(\emptyset,(ij)^c)].
\]
\end{proof}

Based on this, we obtain the following result.
\begin{proposition}
For any 2-additive (normalized) bi-capacity,
\begin{itemize}
\item[(i)] ${\displaystyle \sum_{i\in N}[I_{i,\emptyset}+I_{\emptyset,i}]=2}$.
\item[(ii)] ${\displaystyle \sum_{i\in N}I_{\emptyset,i} =
\frac{1}{2}\sum_{i,j\in N}I_{i,j}+1}$.
\item [(iii)] ${\displaystyle I_{\emptyset,\emptyset} =
-\frac{1}{6}\Big[\sum_{i,j\in N}I_{i,j} + \sum_{ij\subseteq N}[I_{ij,\emptyset} + I_{\emptyset,ij}]\Big]}$.
\end{itemize}
\end{proposition}
\begin{proof} (i)
Since $\sum_{(A,B)\in\Q(N)}m(A,B) =1$, we have:
\begin{align*}
1 &= \sum_{ij\subseteq N}m(\emptyset,(ij)^c) + \sum_{ij\subseteq N}m(i, (ij)^c) +
\sum_{ij\subseteq N}m(ij,(ij)^c) + \sum_{i\in N}m(\emptyset,i^c) + \sum_{i\in
  N}m(i,i^c)+m(\emptyset,N) \\
& = \sum_{ij\subseteq N}I_{\emptyset,ij} + \sum_{ij\subseteq N}I_{i,j} +
\sum_{ij\subseteq N}I_{ij,\emptyset} +\sum_{i\in N}I_{\emptyset,i}
-\frac{1}{2}\sum_{ij\subseteq N}[I_{j,i}+2I_{\emptyset,ij}] +\\
& \sum_{i\in
  N}I_{i,\emptyset} -\frac{1}{2}\sum_{ij\subseteq N}[2I_{ij,\emptyset}+I_{i,j}]-1\\
 &= \sum_{i\in N}I_{i,\emptyset} + \sum_{i\in N}I_{\emptyset,i}-1
\end{align*}
hence the result.

(ii) This is due to $\sum_{B\subsetneq N}m(\emptyset,B) = 1$.

(iii) Using Eq. (\ref{eq:biint}), we have:
\begin{align*}
I_{\emptyset,\emptyset} &= \sum_{(S',T')\in
[(\emptyset,N),(N,\emptyset)]}\frac{1}{n-t'+1}m(S',T')\\
 &=  m(\emptyset,N)+\sum_{(S',T')>(\emptyset,N)}\frac{1}{n-t'+1}m(S',T')\\
 &=  -1 +\frac{1}{3}\sum_{ij\subseteq N}[m(ij,(ij)^c) + m(i,(ij)^c) +
m(\emptyset,(ij)^c)]\\
        & + \frac{1}{2}\sum_{i\in N}[m(i,i^c) + m(\emptyset,i^c)]\\
 &= -1 +\frac{1}{3}\sum_{ij\subseteq N}[I_{ij,\emptyset} + I_{\emptyset,ij} +
I_{i,j}]\\   
        & + \frac{1}{2}\sum_{i\in N}\Big[I_{i,\emptyset} -\frac{1}{2}\sum_{j\neq
i}[I_{i,j} +I_{ij,\emptyset}]+I_{\emptyset,i} - \frac{1}{2}\sum_{j\neq i}[I_{j,i} +
I_{\emptyset,ij}]\Big]\\
 &= -1 -\frac{1}{6}\sum_{ij\subseteq N}[I_{ij,\emptyset}+I_{\emptyset,ij}] -
\frac{1}{6}\sum_{i,j\in N}I_{i,j} +\frac{1}{2}\sum_{i\in N}[I_{i,\emptyset} +
I_{\emptyset,i}] 
\end{align*}
hence the result, using (i) above. 
\end{proof}

We are now able to express the monotonicity conditions for $m$ and $I$. 
\begin{proposition}
Let $m$ be a function defined on $\Q(N)$. It corresponds to the
M\"obius transform  of a normalized bi-capacity $v$ if and only if:
\begin{itemize}
\item [(i)] ${\displaystyle \sum_{(A,B)\in\Q(N)}m(A,B) = 1, \sum_{B\subsetneq
  N}m(\emptyset,B)=1}$, and $m(\emptyset,N)=-1$.
\item [(ii)] For all $i\in N$, for all $(A,B)\in \Q(N\setminus i)$,
  \begin{itemize}
  \item [(ii.1)] ${\displaystyle m(i,i^c) + \sum_{j\in (B\cup i)^c}m(i,(ij)^c) +
  \sum_{j\in A}m(ij,(ij)^c)}\geq 0$
  \item [(ii.2)] ${\displaystyle m(\emptyset,i^c) + \sum_{j\in (B\cup
  i)^c}m(\emptyset, (ij)^c) + \sum_{j\in A}m(j,(ij)^c)}\geq 0$
  \end{itemize}
\end{itemize}  
\end{proposition}
\begin{proof}
Condition (i) is clear. Conditions (ii) come from Lemma 1 of Part I:
\begin{align*}
v(A\cup i,B) - v(A,B) &= \Delta_{(i,i^c)}v(A,B) =
\sum_{(S,T)\in[(i,i^c),(A\cup i,B)]}m(S,T)\geq 0\\
v(A,B) - v(A,B\cup i)&  = \Delta_{(\emptyset,i^c)}v(A,B\cup i) = 
\sum_{(S,T)\in[(\emptyset,i^c),(A,B)]}m(S,T)\geq 0. 
\end{align*}
\end{proof} 

\begin{proposition}
\label{prop:biintm}
Let $I$ be a function defined on $\Q(N)$. It corresponds to the
interaction  of a normalized bi-capacity $v$ if and only if:
\begin{itemize}
\item [(i)] ${\displaystyle \sum_{i\in N} [I_{i,\emptyset} + I_{\emptyset,i}] =
2}$, ${\displaystyle \sum_{i\in N}I_{\emptyset,i} = \frac{1}{2}\sum_{i,j\in
N}I_{i,j}+1}$,\\ ${\displaystyle I_{\emptyset,\emptyset} =
-\frac{1}{6}\Big[\sum_{i,j\in N}I_{i,j} + \sum_{ij\subseteq N}[I_{ij,\emptyset} +
I_{\emptyset,ij}]\Big]}$.
\item [(ii)] For all $i\in N$, for all $(A,B)\in \Q(N\setminus i)$,
  \begin{itemize}
  \item [(ii.1)] \mbox{}
\[
I_{i,\emptyset} + \frac{1}{2}\Big[\sum_{j\in A}I_{ij,\emptyset} -\sum_{j\in (A\cup
    i)^c} I_{ij,\emptyset} - \sum_{j\in B}I_{i,j} + \sum_{j\in (B\cup i)^c}I_{i,j}
\Big]\geq 0 
\]
  \item [(ii.2)] \mbox{}
\[
I_{\emptyset,i} + \frac{1}{2}\Big[\sum_{j\in(B\cup i)^c}I_{\emptyset,ij}
  -\sum_{j\in B} I_{\emptyset,ij} - \sum_{j\in (A\cup i)^c}I_{j,i} + \sum_{j\in
    A}I_{j,i} \Big]\geq 0
\]
  \end{itemize}
\end{itemize}
\end{proposition}
The proof is immediate from previous propositions. 

Let us express the Choquet integral in terms of the M\"obius transform, applying
Prop. \ref{prop:chomo}. With the 2-additive case, we obtain:
\begin{align}
\label{eq:cho2m}
\cho_v(f) & = \sum_{i\in N^-}m(\emptyset,i^c)f_i + \sum_{ij\subseteq
  N^-}m(\emptyset,(ij)^c)(f_i\wedge f_j) + \sum_{\substack{i\in N^-\\j\in
  N^+}}m(\emptyset,(ij)^c)f_i 
  + \nonumber\\
& \sum_{\substack{i\in N^+\\j\in N^-}}m(i,(ij)^c)[(f_i+f_j)\vee 0] +
  \sum_{ij\subseteq N^+}m(i,(ij)^c)f_i + \nonumber\\
& \sum_{ij\subseteq N^+}m(ij,(ij)^c)(f_i\wedge f_j) + \sum_{i\in N^+}
  m(i,i^c)f_i. 
\end{align}
Based on this formula, we obtain the expression in terms of interaction
indices. We propose two formulas.
\begin{proposition}
For any 2-additive $v$, the Choquet integral is given by:
\begin{align}
\label{eq:cho2i}
\cho_v(f) =& \sum_{i\in N^+}f_i\Big[I_{i,\emptyset} + \frac{1}{2}\Big(\sum_{j\in
   N^+\setminus i}I_{i,j} -\sum_{j\in N^-}I_{i,j}-\sum_{j\neq i} I_{ij,\emptyset}\Big)\Big] \nonumber\\
   & + \sum_{i\in N^-}f_i\Big[I_{\emptyset,i} + \frac{1}{2}\Big(-\sum_{j\neq
    i} I_{j,i} +\sum_{j\in N^+}I_{\emptyset,ij} -
   \sum_{j\in N^-\setminus i}I_{\emptyset,ij}\Big)\Big] 
    \nonumber \\
   & + \sum_{ij\subseteq N^-}I_{\emptyset, ij}(f_i\wedge f_j) + \sum_{ij\subseteq
    N^+}I_{ij,\emptyset}(f_i\wedge f_j) + \sum_{\substack{i\in N^+\\ j\in
    N^-}}I_{i,j} [(f_i+f_j)\vee 0].
\end{align}
Another expression is:
\begin{align}
\label{eq:cho2i2}
\cho_v(f) =& \sum_{\substack{ ij\subseteq N^+\\I_{ij,\emptyset}>0}}
  I_{ij,\emptyset}(f_i\wedge f_j) + \sum_{\substack{ij\subseteq N^+ \\
  I_{ij,\emptyset}<0}} |I_{ij,\emptyset}|(f_i\vee f_j)  \nonumber\\
  & +\sum_{\substack {ij\subseteq N^-\\I_{\emptyset,ij}>0}}
  I_{\emptyset,ij}(f_i\wedge f_j) + \sum_{\substack{ij\subseteq N^-\\
  I_{\emptyset,ij}<0}} |I_{\emptyset,ij}|(f_i\vee f_j)  \nonumber\\
  & +\sum_{i\in N^+}f_i\Big[I_{i,\emptyset}+\frac{1}{2}\Big(\sum_{j\in
  N^+\setminus i}\big(I_{i,j} -|I_{ij,\emptyset}|\big) - \sum_{j\in N^-}\big(|I_{i,j}|+I_{ij,\emptyset}\big)
  \Big)\Big] 
  \nonumber\\ 
  &+ \sum_{i\in N^-}f_i\Big[I_{\emptyset,i}+\frac{1}{2}\Big(\sum_{j\in N^+}\big(I_{\emptyset,ij}-|I_{j,i}|\big) -
  \sum_{j\in N^-\setminus i}\big(I_{j,i} +|I_{\emptyset,ij}|\big)\Big)\Big] \nonumber  \\ 
  & - \sum_{\substack{i\in N^+\\j\in N^-\\I_{i,j}<0}}|I_{i,j}|[(f_i+f_j)\vee 0] -
  \sum_{\substack{i\in N^+\\j\in N^-\\I_{i,j}>0}}I_{i,j}[(f_i+f_j)\wedge 0]. 
\end{align}
In addition, the coefficients in $\sum_{i\in N^+}$, $\sum_{i\in N^-}$ are non
negative, and satisfy (partial convexity) 
\begin{align}
\sum_{\substack{ij\subseteq N\\I_{ij,\emptyset}>0}}I_{ij,\emptyset}+
\sum_{\substack{ij\subseteq N\\I_{ij,\emptyset}<0}}|I_{ij,\emptyset}|+ \sum_{i\in
N}\Big[I_{i,\emptyset} - \frac{1}{2}\sum_{j\neq i}\big(I_{i,j} +
|I_{ij,\emptyset}|\big)\Big] &= 1\\
\sum_{\substack{ij\subseteq N\\I_{\emptyset,ij}>0}}I_{\emptyset,ij}+
\sum_{\substack{ij\subseteq N\\I_{\emptyset,ij}<0}}|I_{\emptyset,ij}|+ \sum_{i\in
N}\Big[I_{\emptyset,i} - \frac{1}{2}\sum_{j\neq i}\big(I_{j,i} +
|I_{\emptyset,ij}|\big)\Big] &= 1  
\end{align}
\end{proposition}
\begin{proof}
Eq. (\ref{eq:cho2i}) comes readily from (\ref{eq:cho2m}) and Prop.
\ref{prop:2mI}. In order to show Eq. (\ref{eq:cho2i2}), we start with
(\ref{eq:cho2i}).  Let us express the different terms in the
summation. Remarking that for any $a,b\in \mathbb{R}$, $a\wedge b + a\vee b =
a+b$, we have:
\begin{eqnarray*}
\lefteqn{\sum_{ij\subseteq N^+}I_{ij,\emptyset}(f_i\wedge f_j)=} \\
&& \sum_{\substack{ij\subseteq N^+\\ I_{ij,\emptyset}>0}}I_{ij,\emptyset}(f_i\wedge
f_j) + \sum_{\substack{ij\subseteq N^+\\
    I_{ij,\emptyset}<0}}(-I_{ij,\emptyset})(f_i\vee f_j) -
\sum_{i\in N^+}f_i\sum_{\substack{j\in N^+\setminus i\\
I_{ij,\emptyset}<0}}(-I_{ij,\emptyset}). 
\end{eqnarray*}
Let us regroup this with the term $\sum_{i\in N^+}$. We get:
\begin{align*}
& \sum_{\substack{ij\subseteq N^+\\ I_{ij,\emptyset}>0}}I_{ij,\emptyset}(f_i\wedge
f_j) + \sum_{\substack{ij\subseteq N^+\\
    I_{ij,\emptyset}<0}}(-I_{ij,\emptyset})(f_i\vee f_j) \\
& +  \sum_{i\in N^+}f_i\Big[I_{i,\emptyset} - \frac{1}{2}\sum_{j\in
    N^+}|I_{ij,\emptyset}| - \frac{1}{2}\sum_{j\in N^-}I_{ij,\emptyset}
  -\frac{1}{2} \sum_{j\in N^-}I_{i,j} + \frac{1}{2}\sum_{j\in N^+\setminus
    i}I_{i,j}\Big]. 
\end{align*}
Similarly, we have:
\begin{eqnarray*}
\lefteqn{\sum_{ij\subseteq N^-}I_{\emptyset,ij}(f_i\wedge f_j)=} \\
&& \sum_{\substack{ij\subseteq N^-\\ I_{\emptyset,ij}>0}}I_{\emptyset,ij}(f_i\wedge
f_j) + \sum_{\substack{ij\subseteq N^-\\
    I_{\emptyset,ij}<0}}(-I_{\emptyset,ij})(f_i\vee f_j) -
\sum_{i\in N^-}f_i\sum_{\substack{j\in N^-\setminus i\\
I_{\emptyset,ij}<0}}(-I_{\emptyset,ij}), 
\end{eqnarray*}
and regrouping with the term $\sum_{i\in N^-}$, we get:
\begin{align*}
& \sum_{\substack{ij\subseteq N^-\\ I_{\emptyset,ij}>0}}I_{\emptyset,ij}(f_i\wedge
f_j) + \sum_{\substack{ij\subseteq N^-\\
    I_{\emptyset,ij}<0}}(-I_{\emptyset,ij})(f_i\vee f_j) \\
& +  \sum_{i\in N^-}f_i\Big[I_{\emptyset,i} -\frac{1}{2}\sum_{j\neq i}I_{j,i}
+\frac{1}{2}\sum_{j\in N^+}I_{\emptyset, ij} - \frac{1}{2}\sum_{j\in
N^-\setminus i}|I_{\emptyset,ij}|\Big]. 
\end{align*}
Now, remark that for any $a,b\in \mathbb{R}$, we have $(a+b)\vee 0 +
(a+b)\wedge 0 = a+b$. Applying this, we get:
\begin{eqnarray*}
\lefteqn{\sum_{\substack{i\in N^+\\j\in N^-}}I_{i,j}[(f_i+f_j)\vee 0]=}\\
&& \sum_{\substack{i\in N^+\\j\in N^-\\I_{i,j}<0}}I_{i,j}[(f_i+f_j)\vee 0] +
\sum_{\substack{i\in N^+\\j\in N^-\\I_{i,j}>0}}(-I_{i,j})[(f_i+f_j)\wedge 0]\\
&& - \sum_{i\in N^+}f_i\sum_{\substack{j\in N^-\\I_{i,j}<0}}(-I_{i,j}) -
\sum_{j\in N^-}f_j\sum_{\substack{i\in N^+\\I_{i,j}<0}}(-I_{i,j}). 
\end{eqnarray*}
Regrouping all terms gives the desired result.

We study the signs of the coefficients for $\sum_{i\in N^+}$ and $\sum_{i\in
N^-}$, by applying Prop. \ref{prop:biintm}. For the summation on $N^+$ and a
given $i\in N^+$, taking $A_i:=N^+\cap\{j|I_{ij,\emptyset}<0\}$ and
$B_i:=N^-\cap \{j|I_{i,j}>0\}$, and replacing into (ii.1) of
Prop. \ref{prop:biintm}, we recover the coefficient of $f_i$, which proves that
it is non negative. Similarly for the summation on $N^-$, taking
$A_i:=N^+\cap\{j|I_{j,i}<0\}$, and $B_i:=N^-\cap \{j|I_{\emptyset,ij}>0\}$ proves the non
negativity of the coefficient of $f_i$.

Partial convexity is obtained by letting $f_i=1$ and $f_i=-1$ for all $i$.  
\end{proof}

Although complicated, the second expression gives in detail all interaction
phenomena occuring in a 2-additive bi-capacity model. As for (\ref{eq:icpt}),
the interest of the expression lies in the fact that we have (almost) a convex
sum, so that the importance of each term in the summation is clear. Let us write
down the case of the CPT model. Using Prop. 6 of Part I and
(\ref{eq:conjint}) and denoting $I^+,I^-$ the interactions of $\nu_+,\nu_-$,  we
obtain 
\begin{align*}
\mathrm{CPT}_{\nu^+,\nu^-}(f) =& \sum_{\substack{ ij\subseteq N^+\\I^+(ij)>0}}
  I^+(ij)(f_i\wedge f_j) + \sum_{\substack{ij\subseteq N^+ \\
  I^+(ij)<0}} |I^+(ij)|(f_i\vee f_j) \\
 &+\sum_{\substack{ ij\subseteq N^-\\I^-(ij)>0}}
  I^-(ij)(f_i\vee f_j) + \sum_{\substack{ij\subseteq N^- \\
  I^-(ij)<0}} |I^-(ij)|(f_i\wedge f_j)\\
 &+\sum_{i\in N^+}f_i\Big[I^+(i)-\frac{1}{2}\Big(\sum_{j\in
  N^+\setminus i}|I^+(ij)| + \sum_{j\in N^-}I^+(ij)\Big)\Big] \\
 &+ \sum_{i\in N^-}f_i\Big[I^-(i)-\frac{1}{2}\Big(\sum_{j\in
  N^+}I^-(ij) - \sum_{j\in N^-\setminus i}|I^-(ij)|\Big)\Big].
\end{align*}
If $\nu^+=\nu^-$, then after some rearrangements, we find the symmetric Choquet
integral (\ref{eq:icpt}), or the asymmetric one if $\nu_+=\overline{\nu_-}$.  

\bibliographystyle{plain}
\bibliography{../BIB/fuzzy,../BIB/grabisch,../BIB/general}

\end{document}